\documentclass[journal]{IEEEtran}
\usepackage{cite,amsmath,amssymb,pifont}
\usepackage{subfig}
\usepackage{color}
\usepackage{graphicx}
\usepackage{pstool}
\usepackage{array,booktabs,multirow}

\newcommand{\ve}[1]{\boldsymbol{#1}}
\newcommand{\te}[1]{\overline{\overline{#1}}}

\newcommand{\cmark}{\ding{51}}%
\newcommand{\xmark}{\ding{55}}%

\newcolumntype{N}{@{}m{0pt}@{}}

\newcounter{tempEquationCounter}
\newcounter{thisEquationNumber}
\newenvironment{floatEq}
{\setcounter{thisEquationNumber}{\value{equation}}\addtocounter{equation}{1}
\begin{figure*}[!t]
\normalsize\setcounter{tempEquationCounter}{\value{equation}}
\setcounter{equation}{\value{thisEquationNumber}}
}
{\setcounter{equation}{\value{tempEquationCounter}}
\hrulefill\vspace*{4pt}
\end{figure*}
}

\begin{document}

\title{Influence of Reciprocity, Symmetry and Normal Polarizations on the Angular Scattering Properties of Bianisotropic Metasurfaces}

\author{Karim~Achouri, and Olivier J. F. Martin
}
%

\maketitle

\begin{abstract}
This paper aims at studying the angular scattering properties of bianisotropic metasurfaces and at clarifying the different roles played by tangential and normal polarization densities. Different types of metasurfaces are considered for this study and are classified according to their symmetrical/asymmetrical and reciprocal/nonreciprocal angular scattering behavior. Finally, the paper presents the relationships between the symmetrical angular scattering properties of reciprocal metasurfaces and the structural symmetries of their scattering particles. This may prove to be practically useful for the implementation of metasurfaces with complex angular scattering characteristics.
\end{abstract}


\IEEEpeerreviewmaketitle


\section{Introduction}

Metasurfaces are electrically thin surfaces engineered to control the propagation of electromagnetic waves. In the past few years, they have attracted major attention due to their unprecedented and unmatched capabilities in manipulating light~\cite{yu2014flat,GrbicLightBending,Glybovski20161,Minovich2015,achouri2018design}. They are typically composed of a periodic array of sub-wavelength scattering particles designed to provide a specified scattering response.

In order to model, simulate and implement these structures, several metasurface synthesis and analysis techniques have been developed based on different approaches. For instance, metasurfaces have been modelled based on impedance/admittance matrices~\cite{Pfeiffer2013Huygens,selvanayagam2013circuit}, susceptibility tensors~\cite{achouri2018design,Achouri2015c,achouri2014general} and polarizability tensors~\cite{6805160,6477089}. These techniques have generally in common the concept of modelling metasurfaces as zero-thickness sheets exhibiting effective material parameters. In addition, they also typically ignore the presence of normal polarizations (or currents) with respect to the metasurface plane. The rationale being that normal polarizations may be ignored since electromagnetic fields can be expressed solely in terms of their tangential field components according to the Huygens principle~\cite{kong1986electromagnetic}. The legitimate question of whether normal polarizations are useful in bringing new functionalities to metasurfaces, or could they be simply ignored and replaced by purely tangential polarizations, was then raised~\cite{TretConf,TretEquivalent}. As will be succinctly explained thereafter, normal polarizations may indeed by ignored \emph{if} a metasurface is always excited with the same illumination conditions, e.g. same incidence angle. However, if the incidence angle is changed, then the presence of normal polarizations do play a role in the scattering behavior of metasurfaces and should not be ignored. Additionally, normal polarizations may even be leveraged to bring about new functionalities as demonstrated in~\cite{AngularScattering2017}.

This paper aims at studying the angular scattering behavior of bianisotropic metasurfaces with both tangential and normal polarization densities and clarify the role played by the latter. In particular, we will see how different susceptibility components affect the angular scattering of metasurfaces. We will also discuss how the metasurface scattering particles structural symmetries may be related to the presence of certain susceptibility components. This cannot provide the exact geometry and dimensions of the scattering particles but can at least provide valuable information about the structural symmetries that they should exhibit, which may be of practical interest for designing metasurfaces with complex angular scattering properties.

This paper is organized as follows: Section~\ref{sec:gencon} provides general information regarding the application of the Huygens principle and the various different types of symmetrical/asymmetrical and reciprocal/nonreciprocal scattering properties of metasurfaces. Section~\ref{sec:angscat} presents the angular scattering properties of different types of metasurfaces. Section~\ref{sec:sym} discusses the relationships between the angular scattering symmetries and the scattering particles structural symmetries. Finally, Section~\ref{sec:concl} concludes the paper.

\section{General Considerations}
\label{sec:gencon}

Before delving into the angular scattering properties of metasurfaces and discuss how they may be affected by the structural symmetries of the their scattering particles, we shall first understand how normal polarization densities affect their electromagnetic response and how the latter can be used to bring about new functionalities.

Let us consider the generalized sheet transition conditions (GSTCs), which accurately relate the electromagnetic fields interacting with a metasurface to its material parameters~\cite{kuester2003av,achouri2014general}. Throughout the paper, we will consider that a metasurface is a zero-thickness sheet of polarizable elements lying in the $xy$-plane at $z=0$, we will also consider the time-dependence $e^{j\omega t}$, which we will omit for conciseness. It follows that the GSTCs read
\begin{subequations}
\label{eq:GSTCs}
\begin{equation}\label{eq:GSTCs1}
\hat{z}\times\Delta\ve{H}=j\omega\ve{P} - \hat{z}\times \nabla M_z,
\end{equation}
\begin{equation}\label{eq:GSTCs2}
\hat{z}\times\Delta\ve{E}=-j\omega\mu_0\ve{M} - \frac{1}{\epsilon_0}\hat{z}\times \nabla P_z,
\end{equation}
\end{subequations}
where $\Delta\ve{E}$ and $\Delta\ve{H}$ are respectively the difference of the electric and magnetic fields on both sides of the metasurface, and $\ve{P}$ and $\ve{M}$ are the electric and magnetic polarization densities excited on the metasurface. In the case of bianisotropic metasurfaces, these polarization densities may be expressed in terms of the average electric and magnetic fields on both sides of the metasurfaces as~\cite{kuester2003av}
\begin{subequations}
\label{eq:PM}
\begin{equation}
\ve{P}=\epsilon_0 \te{\chi}_\text{ee}\cdot\ve{E}_\text{av} + \epsilon_0\eta_0 \te{\chi}_\text{em}\cdot\ve{H}_\text{av},
\end{equation}
\begin{equation}
\ve{M}= \te{\chi}_\text{mm}\cdot\ve{H}_\text{av} + \frac{1}{\eta_0} \te{\chi}_\text{me}\cdot\ve{E}_\text{av},
\end{equation}
\end{subequations}
where $\epsilon_0$ is the vacuum permittivity associated with the vacuum impedance, $\eta_0$, and $\te{\chi}_\text{ee}$, $\te{\chi}_\text{mm}$, $\te{\chi}_\text{em}$ and $\te{\chi}_\text{me}$ are the electric, magnetic, electro-magnetic and magneto-electric susceptibility tensors, respectively.

In the most general case, the GSTCs in~\eqref{eq:GSTCs} are differential equations due to the presence of the spatial derivatives of $P_z$ and $M_z$. Moreover, each of the susceptibility tensors in~\eqref{eq:PM} includes both normal and tangential susceptibility components, which amounts to a total of 36 susceptibilities. This makes the modelling of metasurfaces a particularly difficult task. This is one the reasons why it has been common practice to ignore the presence of normal susceptibility components and accordingly discard the spatial derivatives in the GSTCs~\cite{achouri2014general}.

\begin{figure}[h!]
\centering
\includegraphics[width=1\columnwidth]{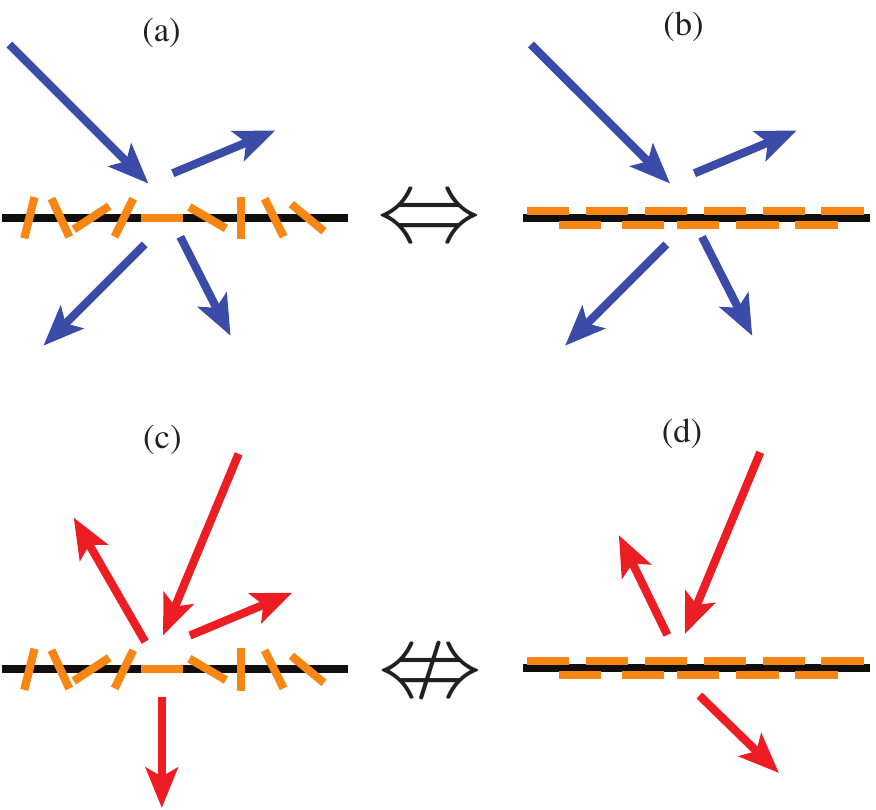}
\caption{Illustrations of the application of the Huygens principle and its limitation. (a) Metasurface with complex orientations of scattering particles in both vertical and  horizontal directions. (b) Metasurface equivalent to (a) but with purely tangential scattering particles, which produces the same scattering as the metasurface in (a). (c) The metasurface in (a) is now excited at an other incidence angle and produces different scattered fields than in (a). (d) The purely tangential metasurface (b) is not the equivalent of the one in (c) anymore.}
\label{fig:Huygens}
\end{figure}

These simplifications are notably justified by three main considerations: 1) in many cases, metasurfaces have been realized for paraxial wave propagation (close to normal incidence) for which the normal components of the electric and magnetic fields are negligible compared to their tangential parts. Therefore, the scattering contributions emerging from the normal polarizations are also negligible. 2) Metasurfaces are very thin compared to the operation wavelength meaning that, irrespectively of the wave propagating angle, the response of the surface is more important in its tangential directions than in its normal direction. 3) The Huygens principle stipulates that an electromagnetic field can always be expressed in terms of its tangential components. Therefore, a metasurface with both normal and tangential susceptibility components can be transformed into an equivalent metasurface with only tangential susceptibility components such that both metasurfaces exhibit the same scattering response~\cite{TretConf,TretEquivalent}, as depicted in Figs.~\ref{fig:Huygens}(a) and~\ref{fig:Huygens}(b). This last point can be easily verified from the GSTCs. Indeed, from~\eqref{eq:GSTCs1}, we see that $\ve{P}$ is on an equal footing with the gradient of $M_z$ and similarly $\ve{M}$ is on an equal footing with the gradient of $P_z$ in~\eqref{eq:GSTCs2}. This means that $M_z$ ($P_z$) can be transformed into an effective and purely tangential electric (magnetic) polarization, $\ve{P}_\text{eff}$ ($\ve{M}_\text{eff}$).

While these three points are generally valid, they also carry their own limitations: 1) metasurfaces can be designed for waves with large propagation angles, making the normal components of their fields actually larger than their corresponding tangential field components. 2) Simple conductive rings or loops within the metasurface plane are sufficient to generate strong normal magnetic responses even in the case of paraxial wave propagation. 3) The application of the Huygens principle to the implementation of metasurfaces with purely tangential polarizations and which exhibit the same scattering response as metasurfaces possessing both normal and tangential polarizations, as depicted in Figs.~\ref{fig:Huygens}a and~\ref{fig:Huygens}b, is of course correct. However, its validity is generally restricted to the case of identical excitations. Indeed, this stems from the presence of the gradients of the polarizations in~\eqref{eq:GSTCs}, which typically limits the validity of the effective purely tangential polarizations, $\ve{P}_\text{eff}$ and $\ve{M}_\text{eff}$, to the case of fixed illumination. It follows that if the illumination angle is changed, then the purely tangential metasurface will, in general, not produce the same scattering as the original metasurface. This is depicted in Figs.~\ref{fig:Huygens}c and~\ref{fig:Huygens}d, where the metasurface in Fig.~\ref{fig:Huygens}c is structurally identical to that in Fig.~\ref{fig:Huygens}a but scatters differently than the metasurface in Fig.~\ref{fig:Huygens}d, which is yet structurally identical to that in Fig.~\ref{fig:Huygens}b.
\begin{figure}[h!]
\centering
\includegraphics[width=0.9\columnwidth]{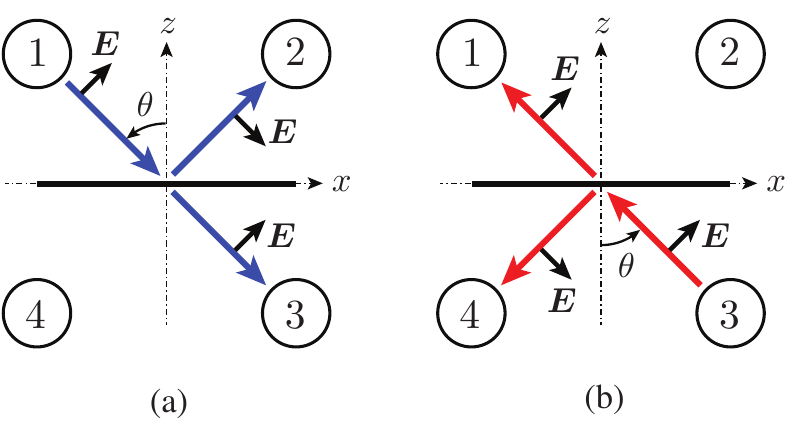}
\caption{Incidence angle and field polarization for (a) downward wave incidence and (b) an upward wave incidence.}
\label{fig:axis}
\end{figure}

\begin{figure*}[h!]
\centering
\includegraphics[width=1\textwidth]{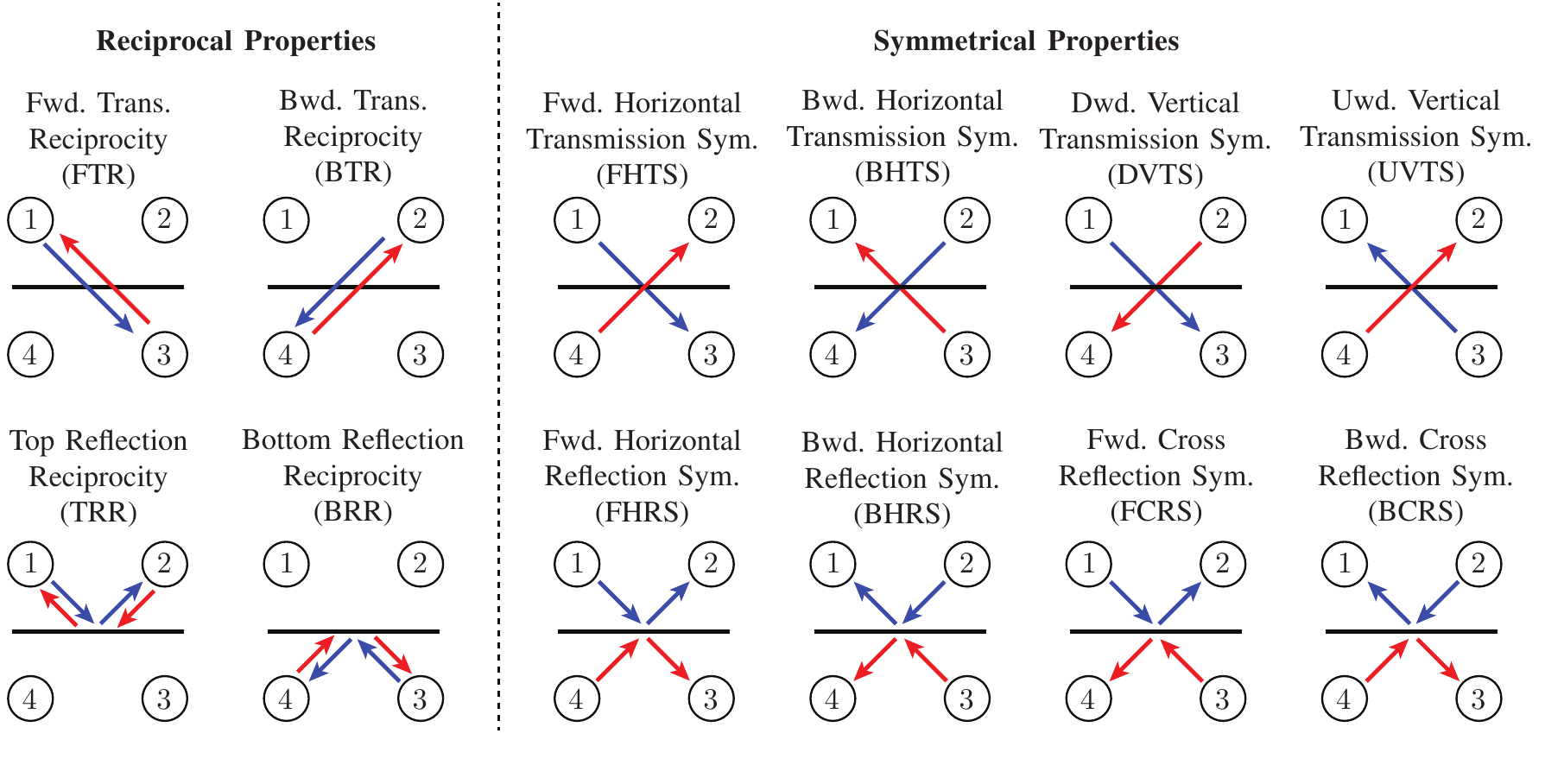}
\caption{Representations of the 12 different possible sorts of scattering. In order they read: Forward Transmission Reciprocity (FTR), Backward Transmission Reciprocity (BTR), Forward Horizontal Transmission Symmetry (FHTS), Backward Horizontal Transmission Symmetry (BHTS), Downward Vertical Transmission Symmetry (DVTS), Upward Vertical Transmission Symmetry (UVTS), Top Reflection Reciprocity (TRR), Bottom Reflection Reciprocity (BRR), Forward Horizontal Reflection Symmetry (FHRS), Backward Horizontal Reflection Symmetry (BHRS), Forward Cross Reflection Symmetry (FCRS), Backward Cross Reflection Symmetry (BCRS).}
\label{fig:Tsym}
\end{figure*}

From these considerations, it follows that the usefulness of the normal polarizations depends upon how a metasurface is used in practice. If it is meant to be illuminated always at the same incidence angle, then the presence of normal polarizations may be ignored. However, if a metasurface is meant to be illuminated at many different incidence angles, then the normal polarizations cannot be ignored and they may even be utilized to bring about additional functionalities, as will be shown thereafter.

Actually, one of the main functionalities that the presence of normal polarizations may add to the already impressive arsenal of field manipulation capabilities of metasurfaces, is their ability in controlling the angular scattering response of the latter, which has been only little studied~\cite{PfeifferEmulating,AngularScattering2017}. In order to understand how the angular scattering response of a metasurface depends upon its susceptibilities, we shall next consider a simplified but yet relevant and pedological scenario.

Let us consider a \emph{uniform} metasurface surrounded by vacuum on both sides and illuminated with a plane wave impinging at an incidence angle $\theta$. This plane wave is reflected and transmitted by the metasurface without rotation of polarization and at the same angle $\theta$ due to the uniformity of the structure. We consider the case where the scattering occurs only in the $xz$-plane and the waves are all p-polarized. The case of s-polarized waves is very similar and is thus not discussed here for briefness. The four quadrants of the $xz$-plane are each associated with a "port", which may serve either as a source or a receiver. The metasurface may then be excited from any of these ports, as illustrated in Fig.~\ref{fig:axis}. It follows that port 1 can only communicate to port 2 via reflection and to port 3 via transmission and can never exchange energy with port 4. And so on for the other ports.

With this approach, we can now easily characterize the angular scattering response of any metasurface. It turns out that there are two main properties that apply in transmission and/or in reflection and which are the properties of \emph{reciprocity} and \emph{symmetry}. In order to understand and visualize what these properties correspond to, one should refer to Fig.~\ref{fig:Tsym} in which we have depicted the 4 different possible types of reciprocal/nonreciprocal scattering and the 8 different possible types of symmetric/asymmetric scattering. We have labelled each of these 12 different cases to be able to easily identify them thereafter.

In the top row of Fig.~\ref{fig:Tsym}, we present the different transmission cases and have thus not drawn the reflected waves for conciseness. Similarly, the bottom row of the figure only represents the different reflection cases for which we have not drawn the transmitted waves.

For the 4 reciprocal cases, the metasurface is either reciprocal, if the blue and red arrows represent the same quantity (e.g. $T_{31}=T_{13}$ and/or $R_{12}=R_{21}$, ...) or nonreciprocal if they are not the same (e.g. $T_{31}\neq T_{13}$ and/or $R_{12}\neq R_{21}$, ...). Similarly, a metasurface would exhibit the property of forward horizontal transmission symmetry (FHTS) if $T_{31} = T_{24}$, and so on.

From Fig.~\ref{fig:Tsym}, one may a priori think that a metasurface has the capability to independently control any of these 12 cases. However, such a feat is not physically possible as some of these properties are actually connected to each other. Indeed, a reciprocal metasurface (one that simultaneously exhibits all 4 reciprocal properties) has all of its transmission/reflection properties which depend on each other, i.e. FHTS $\equiv$ BHTS $\equiv$ DVTS $\equiv$ UVTS and FHRS $\equiv$ BHRS $\equiv$ FCRS $\equiv$ BCRS. Thus, a reciprocal metasurface is either: 1) symmetric in both transmission and reflection, 2) asymmetric in both reflection and transmission, 3) symmetric only in transmission or 4) symmetric only in reflection.

\begin{figure}[h!]
\centering
\includegraphics[width=0.8\columnwidth]{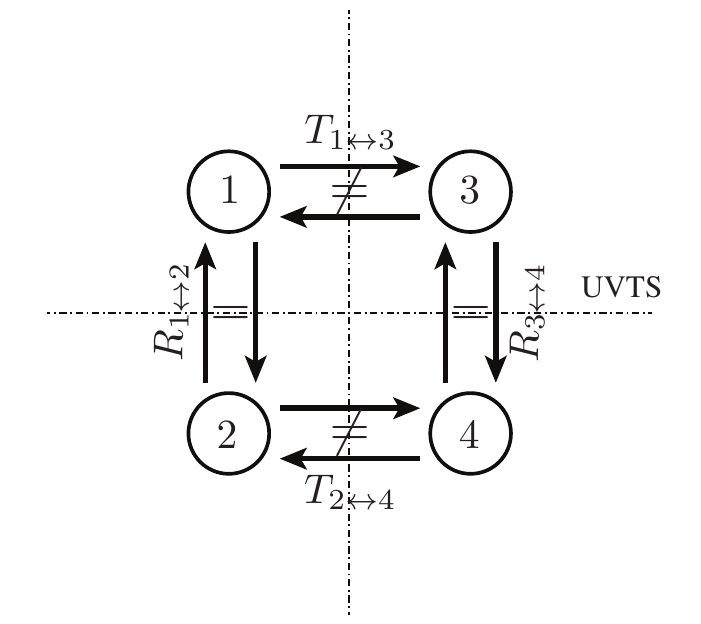}
\caption{Pictorial representation of the scattering properties of a metasurface. The signs $=$ and $\neq$ respectively indicate a reciprocal or a nonreciprocal transmission/reflection relation between two ports. The vertical dashed line indicates that the metasurface exhibits all symmetric reflection properties. The horizontal dashed line with the label UVTS indicates that this metasurface only exhibits this particular transmission symmetry property.}
\label{fig:template}
\end{figure}

Similar relationships exist if only some of the 4 reciprocity conditions are satisfied. For instance, if only the condition of forward transmission reciprocity (FTR) is fulfilled, i.e. $T_{13} = T_{31}$, $T_{24} \neq T_{42}$, $R_{12} \neq R_{21}$ and $R_{34} \neq T_{43}$, then the metasurface can be completely asymmetric in reflection. However, the following transmission symmetry properties would be equivalent to each other: FHTS $\equiv$ UVTS and BHTS $\equiv$ DVTS, thus limiting the capabilities of the metasurface in controlling transmitted waves. This is one example that shows how the concept of nonreciprocity is not just limited to the implementation of isolators but may also be leveraged for additional field manipulations capabilities~\cite{achouri2014general}.

It follows that the symmetric/asymmetric scattering properties of metasurfaces are directly related to their reciprocal/nonreciprocal characteristics. This means that the more nonreciprocal a metasurface is, the more degrees of freedom it generally has to control its angular scattering. As said above, there are 4 types of metasurfaces with different angular scattering properties that could be realized when the 4 reciprocal cases of Fig.~\ref{fig:Tsym} are satisfied. When 3 reciprocal cases are satisfied, then there are 32 different types of metasurfaces that are realizable. If 2 reciprocal cases are satisfied, then 96 different types of metasurfaces can be realized. Finally, if only 1 or none of the reciprocal cases are satisfied, then a total of 256 could be realized.

In order to easily identify the scattering properties of a metasurface, we have introduced the diagrammatic representation depicted in Fig.~\ref{fig:template}, which corresponds to a metasurface with scattering properties arbitrarily chosen for illustration. This type of diagram represents the 4 ports surrounding the metasurface (note placed in the same configuration as in Figs.~\ref{fig:axis} and~\ref{fig:Tsym} for better visualization) and the corresponding transmission and reflection relations between them. A reciprocal relation is indicated with an $=$ sign, while a nonreciprocal relation is indicated with a $\neq$ sign. The vertical dashed line without any label signifies that this metasurface exhibits all reflection symmetries. The horizontal dashed line with the label UVTS indicates that only this transmission symmetry is satisfied. This diagram thus allows one to identify at a glance what are the angular scattering properties of any metasurface.

With all of these considerations in mind, we may now study more specifically the angular scattering properties of metasurfaces in terms of their susceptibilities. This is the topic of the next section.

\section{Angular Scattering Properties of Metasurfaces}
\label{sec:angscat}

In this section, we present the angular transmission and reflection coefficients for 4 different types of metasurfaces. Specifically, we present the case of birefringent metasurfaces with only tangential polarizations, anisotropic metasurfaces with both tangential and normal polarizations, bianisotropic metasurfaces with only tangential polarizations and bianisotropic metasurfaces with tangential and normal polarizations. For each of these cases, we provide numerical simulations of reciprocal metasurfaces built from actual scattering particles.

In order to properly assess the angular scattering behavior of these metasurfaces in terms of their susceptibilities, we derive the expressions of their scattering parameters. To do so, we first have to define the difference and average of the fields in~\eqref{eq:GSTCs} and~\eqref{eq:PM}. Using the convention adopted in Fig.~\ref{fig:axis}, we have at $z=0$ that
\begin{subequations}\label{eq:DiffAvDown}
\begin{equation}\label{eq:DiffAvDown.a}
\Delta\ve{E} = \pm\hat{x}\frac{k_z}{k_0}\left(1 + R -T\right),
\end{equation}
\begin{equation}
\Delta\ve{H} = \hat{y}\frac{1}{\eta_0}\left(-1 + R + T\right),
\end{equation}
\begin{equation}
E_{x,\text{av}} = \frac{k_z}{2k_0}\left(1 + T + R\right),
\end{equation}
\begin{equation}
E_{z,\text{av}} = \frac{k_x}{2k_0}\left(1 + T - R\right),
\end{equation}
\begin{equation}\label{eq:DiffAvDown.e}
H_{y,\text{av}} = \mp\frac{1}{2\eta_0}\left(1 + T - R\right),
\end{equation}
\end{subequations}
where $k_z = k_0\cos\theta$ and $k_x = k_0\sin\theta$ and where we have dropped the term $e^{-jk_xx}$ for conciseness. In~\eqref{eq:DiffAvDown.a} and~\eqref{eq:DiffAvDown.e}, the top signs correspond to incident waves propagating backward along $z$, as in Fig.~\ref{fig:axis}a, while the bottom signs correspond to incident waves propagating forward as in Fig.~\ref{fig:axis}b.

Now, one can obtain the transmission and reflection coefficients of any metasurface simply by substituting~\eqref{eq:DiffAvDown} into~\eqref{eq:GSTCs} and~\eqref{eq:PM} and solving the resulting system of equations for the parameters $T$ and $R$, respectively. The material properties of the metasurface are encoded into its susceptibilities and one can thus decide what type of metasurface to study just by setting to zero the susceptibility components that are not desired. Note that if a metasurface is reciprocal, then some of its susceptibility components are related to each other as stipulated by the Lorentz reciprocity theorem~\cite{kong1986electromagnetic}. Accordingly, the susceptibility tensors of a reciprocal metasurface must satisfy the following reciprocity conditions
\begin{equation}
\label{eq:reciprocity}
\te{\chi}_\text{ee}^\text{T}=\te{\chi}_\text{ee},\qquad
\te{\chi}_\text{mm}^\text{T}=\te{\chi}_\text{mm},\qquad
\te{\chi}_\text{me}^\text{T}=-\te{\chi}_\text{em}.
\end{equation}
A contrario, a nonreciprocal metasurface does not satisfy at least one of these conditions.

\subsection{Birefringent Metasurfaces}
\label{sec:biref}

We now start by considering the case of birefringent metasurfaces, which is the most conventional and simple type of metasurfaces. Assuming p-polarized waves, the only susceptibility components that are excited on such a metasurface are $\chi_\text{ee}^{xx}$ and $\chi_\text{mm}^{yy}$. This metasurface thus does not exhibit any normal polarization and is always fully reciprocal according to~\eqref{eq:reciprocity}. The corresponding angular scattering parameters are given by
\begin{subequations}
\begin{equation}
R = \frac{2j(k_0^2\chi_\text{mm}^{yy}-k_z^2\chi_\text{ee}^{xx})}{(2 + jk_z\chi_\text{ee}^{xx})(2k_z + jk_0^2\chi_\text{mm}^{yy})},
\end{equation}
\begin{equation}
T = \frac{k_z(4 +k_0^2\chi_\text{mm}^{yy}\chi_\text{ee}^{xx})}{(2 + jk_z\chi_\text{ee}^{xx})(2k_z + jk_0^2\chi_\text{mm}^{yy})}.
\end{equation}
\end{subequations}

\begin{figure}[h!]
\centering
\includegraphics[width=0.8\columnwidth]{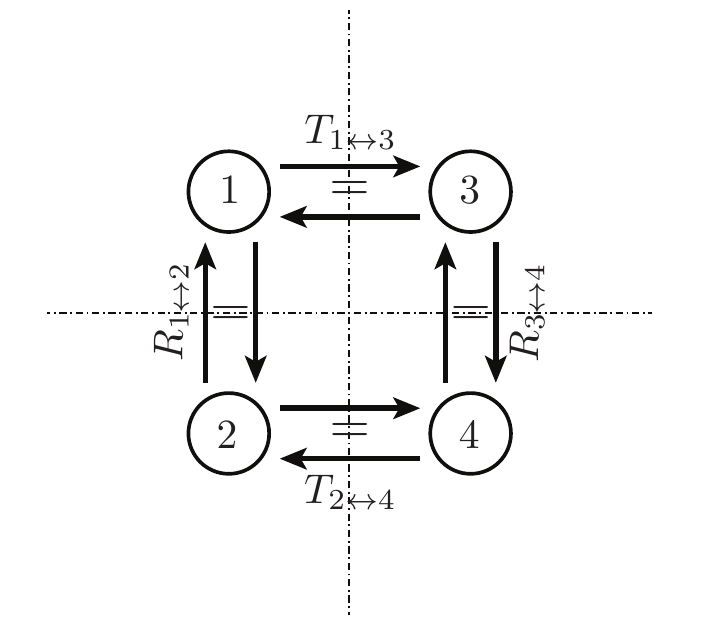}
\caption{Angular scattering properties of birefringent metasurfaces.}
\label{fig:biref}
\end{figure}

We can directly see that these coefficients do not depend on $k_x$ meaning that changing the incidence angle from $\theta$ to $-\theta$ will not change the response of the metasurface. Moreover, illuminating the metasurface from the top or from the bottom does not affect its response either. We conclude that such a metasurface is fully symmetrical in addition of being fully reciprocal. Its corresponding diagrammatical representation is thus that of Fig.~\ref{fig:biref}.

We now present a realization of such a birefringent metasurface. Let us consider the unit cell structure of Fig.~\ref{fig:All}a, which composes the metasurface with a square lattice period of 200~nm. The unit cell is composed of two identical gold rods separated by 50~nm. They have a square cross section of $40\times40$~nm and a length of 170~nm. This unit cell structure is thus perfectly symmetric in the $x$-, $y$- and $z$-directions, which results in the expected fully symmetrical behavior. The resulting transmission (solid lines) and reflection (dashed lines) amplitude and phase, computed using a full-wave commercial software, are plotted in Figs.~\ref{fig:All}(e) and~\ref{fig:All}(i), respectively.
%
%
As can be verified in these plots, this metasurface is indeed completely symmetrical.

\subsection{Anisotropic Metasurfaces}

We now discuss the more general case of anisotropic metasurfaces. For the considered case of p-polarization, the only susceptibilities excited on such metasurfaces are $\chi_\text{ee}^{xx}$, $\chi_\text{ee}^{xz}$, $\chi_\text{ee}^{zx}$, $\chi_\text{ee}^{zz}$ and $\chi_\text{mm}^{yy}$. The resulting scattering parameters are given by
\begin{subequations}
\label{eq:aniTR}
\begin{equation}\label{eq:aniTR1}
R = \frac{2}{C_1}\left[k_0^2\chi_\text{mm}^{yy}-k_z^2\chi_\text{ee}^{xx}+k_xk_z(\chi_\text{ee}^{zx}-\chi_\text{ee}^{xz})+k_x^2\chi_\text{ee}^{zz}\right],
\end{equation}
\begin{equation}\label{eq:aniTR2}
\begin{split}
T = & \frac{jk_z}{C_1}\big[2jk_x(\chi_\text{ee}^{xz}+\chi_\text{ee}^{zx})+k_x^2(\chi_\text{ee}^{xz}\chi_\text{ee}^{zx}-\chi_\text{ee}^{xx}\chi_\text{ee}^{zz})\\
&\qquad\quad -4-k_0^2\chi_\text{ee}^{xx}\chi_\text{mm}^{yy})\big],
\end{split}
\end{equation}
\begin{equation}
\begin{split}
&C_1 = 2(k_z^2 \chi_\text{ee}^{xx}+k_x^2\chi_\text{ee}^{zz}+k_0^2\chi_\text{mm}^{yy})\\
&\qquad +jk_z\left[k_x^2(\chi_\text{ee}^{xx}\chi_\text{ee}^{zz}-\chi_\text{ee}^{xz}\chi_\text{ee}^{zx})-4+k_0^2\chi_\text{ee}^{xx}\chi_\text{mm}^{yy}\right],
\end{split}
\end{equation}
\end{subequations}
%
%
%
where $T$ and $R$ are the same whether the metasurface is illuminated from the top or from the bottom. This already provides an important information, which is that this type of metasurfaces always satisfies \emph{at least} the conditions of FCRS and BCRS. In addition, we also see the presence of $k_x$ for both the transmission and the reflection coefficients. We can thus infer that this type of metasurfaces may present some form of asymmetric scattering in both transmission and reflection. To be more specific, we next consider the cases of reciprocal and nonreciprocal scattering separately.

\begin{figure}[h!]
\centering
\includegraphics[width=0.8\columnwidth]{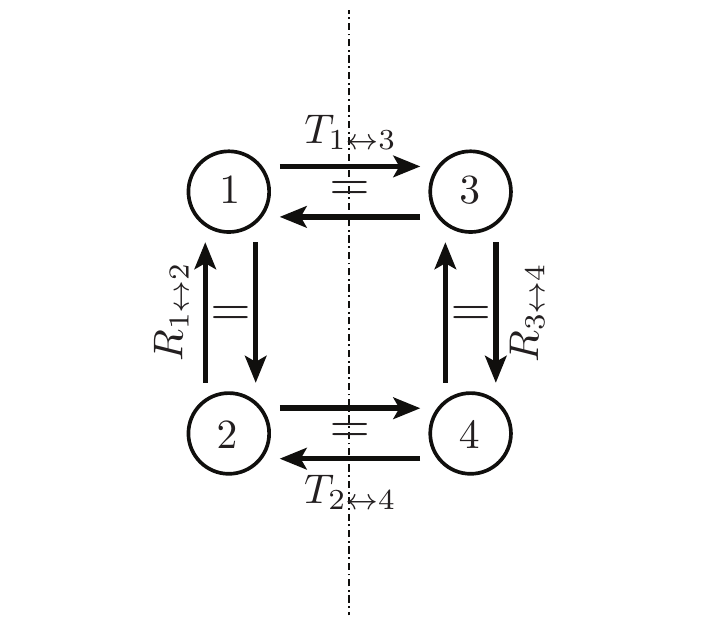}
\caption{Angular scattering properties of reciprocal anisotropic metasurfaces.}
\label{fig:ani_rec}
\end{figure}

Let us first consider the case of a reciprocal metasurface for which $\chi_\text{ee}^{xz}=\chi_\text{ee}^{zx}$ according to~\eqref{eq:reciprocity}. In this case, we see that $k_x$ disappear in~\eqref{eq:aniTR1}, leaving only $k_x^2$ in its numerator and denominator. This confirms the reciprocal behavior of this metasurface as the reflection coefficient remains the same for incidence angles $\theta$ and $-\theta$. In contrast, the coefficient $k_x$ does not disappear in~\eqref{eq:aniTR2} meaning that such a metasurface is transmission asymmetric. As a result, this type of metasurface may be represented by the diagrammatic illustration of Fig.~\ref{fig:ani_rec}.

In order to verify the angular scattering properties of such a metasurface, we have designed the unit cell shown in Fig.~\ref{fig:All}b. It has the same dimensions as the structure in Fig.~\ref{fig:All}a with the addition of two vertical rods with a length of 75~nm. As can be seen in the simulation results of Figs.~\ref{fig:All}f and~\ref{fig:All}j, the metasurface is perfectly symmetric in reflection and exhibits a strong angular asymmetric transmission.

\begin{figure}[h!]
\centering
\includegraphics[width=0.8\columnwidth]{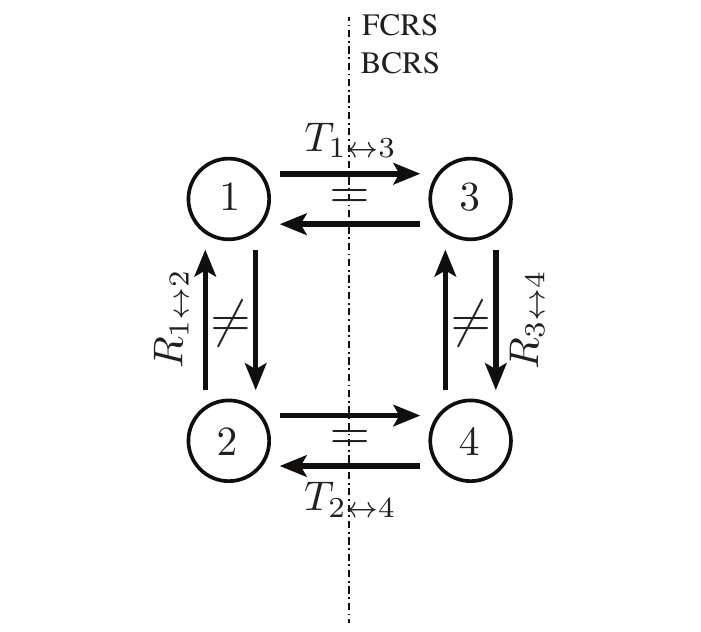}
\caption{Angular scattering properties of nonreciprocal anisotropic metasurfaces.}
\label{fig:ani_NR}
\end{figure}

Let us now consider the case where the metasurface is nonreciprocal, i.e. when $\chi_\text{ee}^{xz}\neq\chi_\text{ee}^{zx}$. Then again, the metasurface is completely transmission asymmetric. Except in the very particular scenario where $\chi_\text{ee}^{xz}=-\chi_\text{ee}^{zx}$, in which case the term $k_x$ vanishes in the nominator of~\eqref{eq:aniTR2} and the metasurface becomes fully symmetric in transmission. Besides that particular case, such a metasurface is only nonreciprocal in reflection, while being fully reciprocal in transmission. Indeed, the term $k_x$ does not vanish in~\eqref{eq:aniTR1} leading to a nonreciprocal reflection coefficient and since the scattering parameters are the same irrespectively of the illumination side, the metasurface is always reciprocal in transmission. In addition, such a metasurface always satisfies the properties of FCRS and BCRS since the reflection coefficient is the same for both sides but different for $\theta$ and $-\theta$. It follows that, for the general case of nonreciprocal scattering (i.e. when $\chi_\text{ee}^{xz}\neq\pm\chi_\text{ee}^{zx}$), the diagrammatic representation of this metasurface is that of Fig~\ref{fig:ani_NR}.

It is important to consider that the asymmetry of this anisotropic metasurface, whether it is reciprocal or not, is directly related to the presence of the term $k_x$ and not that of $k_x^2$ since the latter is a symmetric function of $\theta$. Upon inspection of~\eqref{eq:aniTR}, we see that the presence of $k_x$ is related to the presence of $\chi_\text{ee}^{xz}$ and $\chi_\text{ee}^{zx}$, while $\chi_\text{ee}^{zz}$ is essentially related to $k_x^2$. It follows that the susceptibility components $\chi_\text{ee}^{xz}$ and $\chi_\text{ee}^{zx}$  are responsible for the asymmetric angular scattering of this metasurface. In fact, it would be impossible to distinguish between a metasurface where the only nonzero normal susceptibility is $\chi_\text{ee}^{zz}$ and the birefringent metasurfaces of Sec.~\ref{sec:biref} since $\chi_\text{ee}^{zz}$ does not break the scattering symmetry of the structure. In other words, even if a birefringent metasurface would exhibit a nonzero $\chi_\text{ee}^{zz}$ component, its angular scattering response would remain the same.


\subsection{Bianisotropic Metasurfaces with only Tangential Polarizations}

Let us now consider the case of bianisotropic metasurfaces with only tangential polarizations densities for which, assuming p-polarization, only the following susceptibilities are excited: $\chi_\text{ee}^{xx}$, $\chi_\text{mm}^{yy}$, $\chi_\text{em}^{xy}$ and $\chi_\text{me}^{yx}$. The corresponding scattering parameters are given by
\begin{subequations}
\label{eq:bianiTR}
\begin{equation}\label{eq:bianiTR1}
R = \frac{2k_0^2\chi_\text{mm}^{yy}-2k_z\left[k_z\chi_\text{ee}^{xx}\mp k_0(\chi_\text{me}^{yx}-\chi_\text{em}^{xy})\right]}{k_z\left[2k_z\chi_\text{ee}^{xx}-j(4+k_0^2\chi_\text{em}^{xy}\chi_\text{me}^{yx}) \right] + k_0^2(2+jk_z\chi_\text{ee}^{xx})\chi_\text{mm}^{yy}},
\end{equation}
\begin{equation}\label{eq:bianiTR2}
\begin{split}
T = \frac{jk_z\left[(2j\mp k_0\chi_\text{em}^{xy})(2j\mp k_0\chi_\text{me}^{yx})-k_0^2\chi_\text{ee}^{xx}\chi_\text{mm}^{yy}\right]}{k_z\left[2k_z\chi_\text{ee}^{xx}-j(4+k_0^2\chi_\text{em}^{xy}\chi_\text{me}^{yx}) \right] + k_0^2(2+jk_z\chi_\text{ee}^{xx})\chi_\text{mm}^{yy}},
\end{split}
\end{equation}
\end{subequations}
where the top signs correspond to incident waves propagating backward along $z$, as in Fig.~\ref{fig:axis}a, while the bottom signs correspond to incident waves propagating forward as in Fig.~\ref{fig:axis}b.

\begin{figure}[h!]
\centering
\includegraphics[width=0.8\columnwidth]{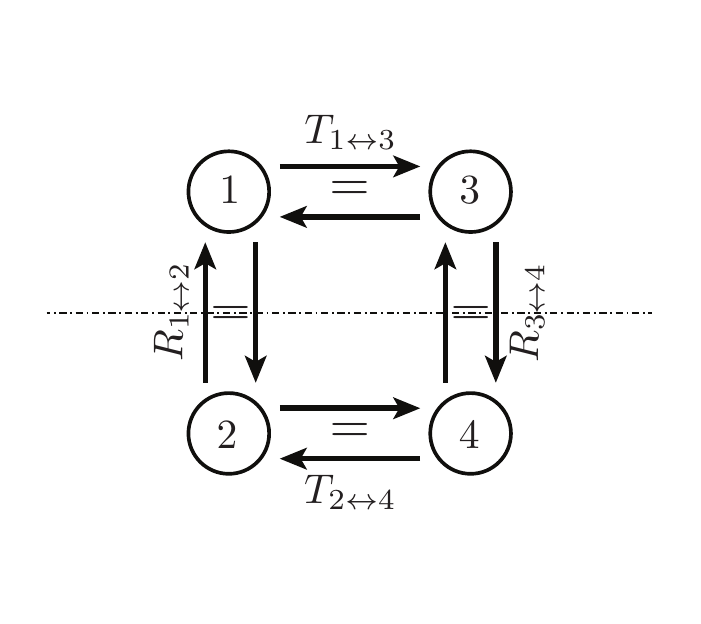}
\caption{Angular scattering properties of reciprocal bianisotropic metasurfaces with only tangential polarizations.}
\label{fig:biani_no_z_R}
\end{figure}

\begin{figure*}[h]
\centering
\includegraphics[width=1\textwidth]{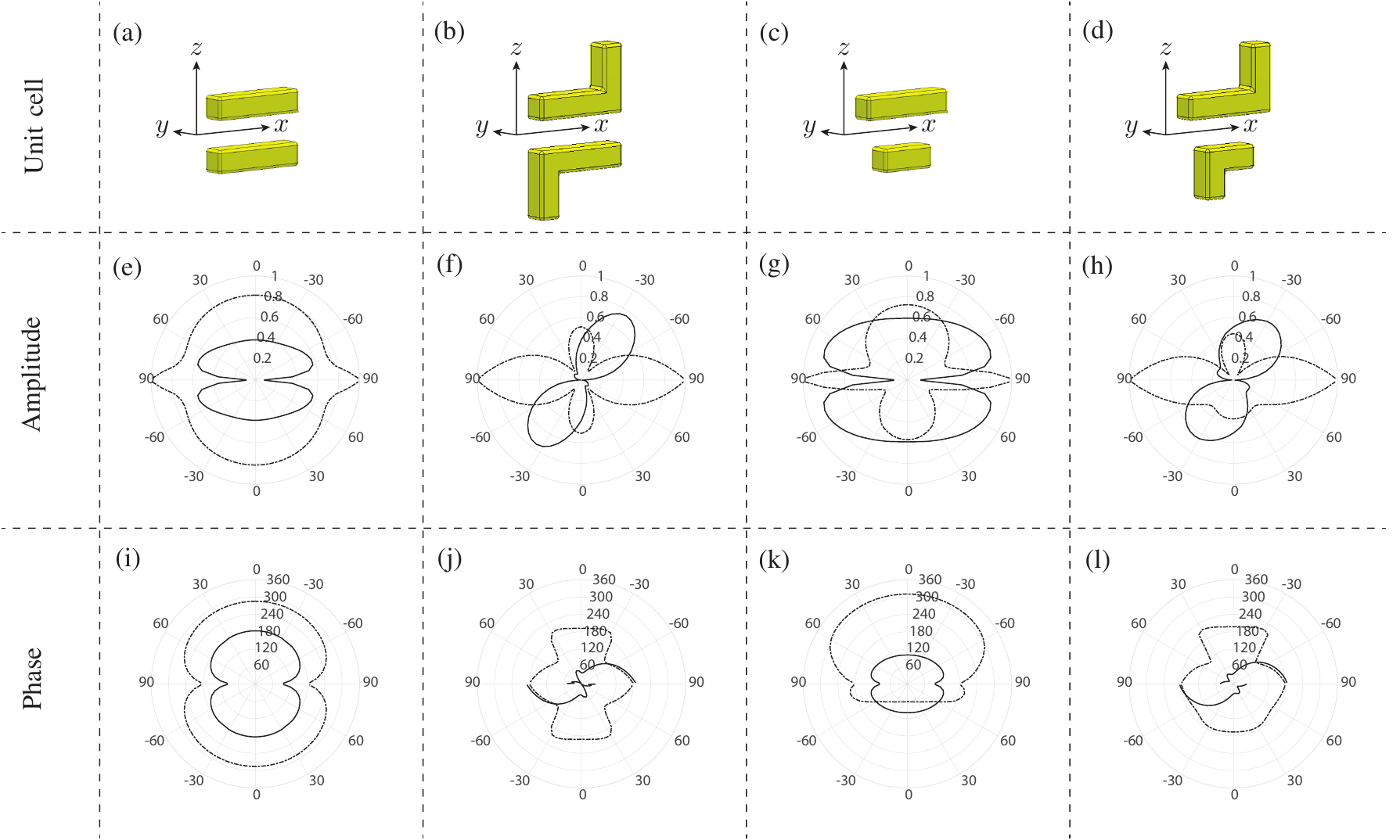}
\caption{Angular scattering properties of 4 different reciprocal metasurfaces. Top row, metasurface unit cells which are periodically arranged in the $xy$-plane with a square lattice period of 200~nm to form the corresponding metasurfaces. Middle row, amplitude of the transmission (solid lines) and reflection (dashed-dotted lines) coefficients versus incidence angle. Note that the angular coordinate of these plots corresponds to the incidence angle $\theta$ following the convention adopted in Figs.~\ref{fig:axis}a and~\ref{fig:axis}b. Bottom row, phase of the transmission and reflection coefficients. The unit cells in (a) and (c) have been simulated at $\lambda_0 = 600$~nm, while the unit cells (b) and (d) have been simulated at $\lambda_0 = 660$~nm.}
\label{fig:All}
\end{figure*}

As can be seen, these expressions do not depend on $k_x$. We can thus directly infer that the scattering response is the same for incidence angles of $\theta$ and $-\theta$. Moreover, the fact that the scattering parameters are not the same for top and bottom illuminations indicates that such a metasurface must exhibit some form of asymmetric scattering. In order to be more specific, we again separate the cases of reciprocal and nonreciprocal scattering.

Let us first discuss the case when the metasurface is reciprocal, i.e. when $\chi_\text{em}^{xy} = - \chi_\text{me}^{yx}$ according to~\eqref{eq:reciprocity}. When this reciprocity condition is satisfied, the transmission coefficient in~\eqref{eq:bianiTR2} becomes the same for top and bottom illuminations. It follows that  the transmission coefficient is fully reciprocal and symmetric. However, the dependence on the illumination side does not vanish for the reflection coefficient in~\eqref{eq:bianiTR1}, meaning that such a metasurface is completely reflection asymmetric. The corresponding diagrammatic representation is depicted in Fig.~\ref{fig:biani_no_z_R}.

\begin{figure}[h!]
\centering
\includegraphics[width=0.8\columnwidth]{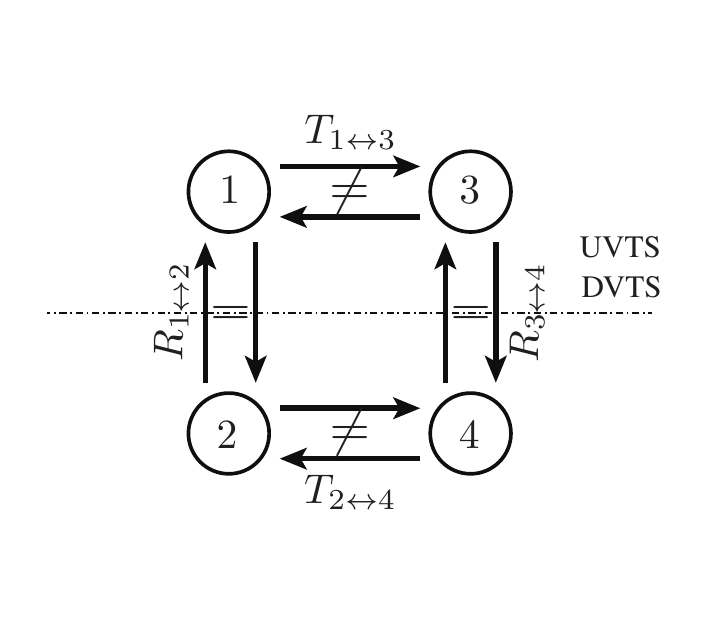}
\caption{Angular scattering properties of nonreciprocal bianisotropic metasurfaces with only tangential polarizations.}
\label{fig:biani_no_z_NR}
\end{figure}

From these considerations, we must conclude that the bianisotropic susceptibilities $\chi_\text{em}^{xy}$ and $\chi_\text{me}^{yx}$  are nonzero when a metasurface presents a structural asymmetry in its normal direction. Concretely, the metasurface does not ``look'' the same when seen from both sides. Such a peculiar property may, for instance, allow one to control the reflection phase or the matching condition depending on the illumination side. This has been notably leveraged in~\cite{8259235} to implement fully efficient refractive metasurfaces, where the incident and refracted beams can both be matched even though they propagate at different angles.

In order to demonstrate the angular scattering properties of this type of metasurfaces, we have designed the unit cell of Fig.~\ref{fig:All}c. It consists of two rods with a respective length of 170~nm and 100~nm. The two rods do not have the same length such that the metasurface is structurally asymmetric with respect to its normal direction. The resulting amplitude and phase of the scattering parameters are respectively plotted in Figs.~\ref{fig:All}g and~\ref{fig:All}k, where we indeed retrieve the expected asymmetric reflection behavior.

Now, let us consider the case where the metasurface is nonreciprocal, i.e. when $\chi_\text{em}^{xy} \neq - \chi_\text{me}^{yx}$. We have that the dependence on the illumination side does not vanish anymore and the metasurface is thus nonreciprocal in transmission. However, the metasurface remains reciprocal in reflection because of the absence of $k_x$ term in~\eqref{eq:bianiTR}. Due to the structural asymmetry and the nonreciprocal transmission properties of such a metasurface, only the transmission symmetries UVTS and DVTS are satisfied. It follows that the diagrammatic representation of such a metasurface is that of Fig.~\ref{fig:biani_no_z_NR}.

Note that there is a special case of nonreciprocity where $\chi_\text{em}^{xy} = \chi_\text{me}^{yx}$. If this equality is satisfied, then the reflection coefficient becomes the same for top and bottom illuminations and the metasurface thus becomes fully reflection symmetric.

Finally, we would like to discuss a very peculiar but particularly interesting case of bianisotropic metasurfaces, which is when $\chi_\text{ee}^{xx} = \chi_\text{mm}^{yy} = 0$ and $\chi_\text{em}^{xy} \neq 0$ and $\chi_\text{me}^{yx}\neq 0$. In such a scenario, the scattering parameters in~\eqref{eq:bianiTR} reduce to
\begin{subequations}
\label{eq:bianiTRred}
\begin{equation}\label{eq:bianiTRred1}
R = \pm\frac{2jk_0(\chi_\text{em}^{xy} - \chi_\text{me}^{yx})}{4+k_0^2\chi_\text{em}^{xy}\chi_\text{me}^{yx}},
\end{equation}
\begin{equation}\label{eq:bianiTRred2}
T = \frac{(2\pm jk_0\chi_\text{em}^{xy})(2\pm jk_0\chi_\text{me}^{yx})}{4+k_0^2\chi_\text{em}^{xy}\chi_\text{me}^{yx}}.
\end{equation}
\end{subequations}
The particular characteristics of such a metasurface is that it does not exhibit any angular dependence, i.e. the coefficients~\eqref{eq:bianiTRred} do not contain $k_x$ or $k_z$. Accordingly, its scattering is the same irrespectively of the incidence angle. One example of such a metasurface is when $\chi_\text{em}^{xy}=-\chi_\text{me}^{yx}=-2j/k_0$. In this case, the resulting metasurface is reciprocal, passive and lossless (according to the corresponding hermitian conditions~\cite{kong1986electromagnetic}) and behaves as a perfect magnetic conductor ($R=1$ and $T=0$) when illuminated from the top and a perfect electric conductor ($R=-1$ and $T=0$) when illuminated from bottom.

\subsection{Bianisotropic Metasurfaces with Tangential and Normal Polarizations}

We shall now consider the most general type of metasurfaces under p-polarized excitation, which is that of bianisotropic metasurfaces with both tangential and normal polarizations. The susceptibility components that may be excited on such metasurfaces are: $\chi_\text{ee}^{xx}$, $\chi_\text{ee}^{xz}$, $\chi_\text{ee}^{zx}$, $\chi_\text{ee}^{zz}$, $\chi_\text{mm}^{yy}$, $\chi_\text{em}^{xy}$, $\chi_\text{me}^{yx}$, $\chi_\text{em}^{zy}$ and $\chi_\text{me}^{yz}$. The corresponding scattering parameters are given in Eqs.~\eqref{eq:RTUHP} below, where, as before, the top signs correspond to illumination from the top and the bottom signs to illumination from the bottom.

\begin{figure}[h!]
\centering
\includegraphics[width=0.8\columnwidth]{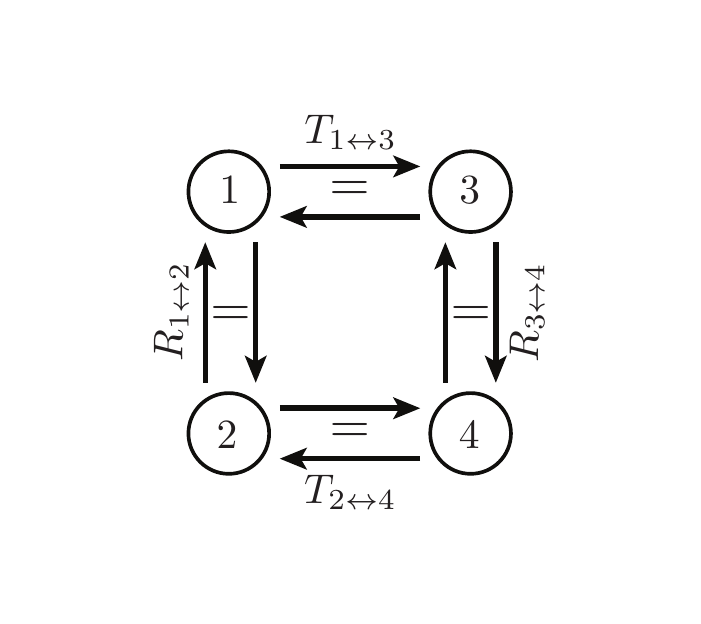}
\caption{Angular scattering properties of reciprocal bianisotropic metasurfaces with tangential and normal polarizations.}
\label{fig:biani_rec}
\end{figure}

As may be expected, combining the effects of $\chi_\text{ee}^{xz}$ and $\chi_\text{ee}^{zx}$, and $\chi_\text{em}^{xy}$ and $\chi_\text{me}^{yx}$, leads to metasurfaces that have the capabilities of being completely asymmetric in both reflection and transmission. As explained before, the presence of $\chi_\text{ee}^{zz}$ is ``neutral'' in the sense that this susceptibility is related to $k_x^2$, which is a symmetric function of $\theta$. The presence of $\chi_\text{em}^{zy}$ and $\chi_\text{me}^{yz}$ plays a role that is similar to that of $\chi_\text{ee}^{xz}$ and $\chi_\text{ee}^{zx}$ since both sets of susceptibilities are related to the presence of $k_x$. However, if the metasurface is reciprocal, then $\chi_\text{em}^{zy}$ and $\chi_\text{me}^{yz}$ cancel each other in~\eqref{eq:RTUHP} (since by reciprocity: $\chi_\text{em}^{zy}=-\chi_\text{me}^{yz}$) and thus do not play a role in the scattering. These two susceptibilities may thus be used as a mean to break the angular scattering symmetries but only in nonreciprocal metasurfaces.

\begin{figure}[h!]
\centering
\includegraphics[width=0.8\columnwidth]{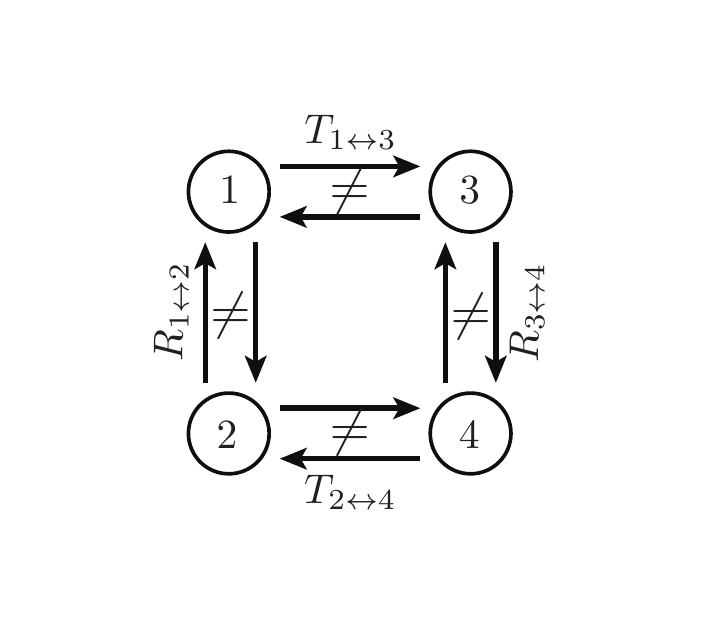}
\caption{Angular scattering properties of nonreciprocal bianisotropic metasurfaces with tangential and normal polarizations.}
\label{fig:biani_NR}
\end{figure}

\begin{subequations}
\label{eq:RTUHP}
\begin{floatEq}
\begin{equation}
R = \frac{2}{C_2}\left\{k_x^2\chi_\text{ee}^{zz}-k_z^2\chi_\text{ee}^{xx}-k_z\left[k_x(\chi_\text{ee}^{xz}-\chi_\text{ee}^{zx})\mp k(\chi_\text{em}^{xy}-\chi_\text{me}^{yx})\right]\mp kk_x(\chi_\text{em}^{zy}+\chi_\text{me}^{yz})+k^2\chi_\text{mm}^{yy} \right\}.
\end{equation}
\begin{equation}
\begin{split}
T = \frac{jk_z}{C_2}\big\{k_x^2(\chi_\text{ee}^{xz}&\chi_\text{ee}^{zx} - \chi_\text{ee}^{xx}\chi_\text{ee}^{zz})+(2j\mp k\chi_\text{em}^{xy})(2j\mp k\chi_\text{me}^{yx})\\
&+k_x\left[\chi_\text{ee}^{zx}(2j\mp k\chi_\text{em}^{xy})+\chi_\text{ee}^{xz}(2j\mp k\chi_\text{me}^{yx})\pm k\chi_\text{ee}^{xx}(\chi_\text{em}^{zy}+\chi_\text{me}^{yz})\right]-k^2\chi_\text{ee}^{xx}\chi_\text{mm}^{yy}\big\}.
\end{split}
\end{equation}
\begin{equation}
\begin{split}
C_2 =  &2\left[k_z^2\chi_\text{ee}^{xx}+k_x^2\chi_\text{ee}^{zz} \mp k k_x (\chi_\text{em}^{zy}+\chi_\text{me}^{yz}) + k^2\chi_\text{mm}^{yy} \right] \pm k^2(\chi_\text{ee}^{xx}\chi_\text{mm}^{yy}-\chi_\text{em}^{xy}\chi_\text{me}^{yx})\\
&\qquad\qquad\quad - jk_z\left[k_x^2(\chi_\text{ee}^{xz}\chi_\text{ee}^{zx}-\chi_\text{ee}^{xx}\chi_\text{ee}^{zz})+4\mp kk_x(\chi_\text{ee}^{zx}\chi_\text{em}^{xy}+\chi_\text{ee}^{xz}\chi_\text{me}^{yx}-\chi_\text{ee}^{xx}(\chi_\text{em}^{zy}+\chi_\text{me}^{yz}))\right].
\end{split}
\end{equation}
\end{floatEq}
\end{subequations}
%
%

\begin{table*}[h!]
\centering
\begin{tabular}{c||c||c|c|c|c|c|c||c|c|c|c|c|cN}
Type  & Reciprocity & FTR & BTR & FHTS & BHTS & DVTS & UVTS & TRR & BRR & FHRS & BHRS & FCRS & BCRS &\\
\hline\hline \begin{tabular}[c]{@{}c@{}} Birefringent\\$\chi_\text{ee}^{xx}$, $\chi_\text{mm}^{yy}$\end{tabular} & \cmark & \cmark & \cmark & \cmark & \cmark & \cmark & \cmark & \cmark & \cmark & \cmark & \cmark & \cmark & \cmark &\\[12pt]
\hline\hline \multirow{3}{*}[-0.5cm]{\begin{tabular}[c]{@{}c@{}c@{}} Anisotropic\\$\chi_\text{ee}^{xx}$, $\chi_\text{mm}^{yy}$\\ $\chi_\text{ee}^{xz}$, $\chi_\text{ee}^{zx}$, $\chi_\text{ee}^{zz}$\end{tabular}} & \begin{tabular}[c]{@{}c@{}} \cmark\\ $\chi_\text{ee}^{xz}=\chi_\text{ee}^{zx}$\end{tabular} & \cmark & \cmark & \xmark & \xmark & \xmark & \xmark & \cmark & \cmark & \cmark & \cmark & \cmark & \cmark &\\[15pt]\cline{2-14}
  & \begin{tabular}[c]{@{}c@{}} \xmark\\ $\chi_\text{ee}^{xz}\neq\chi_\text{ee}^{zx}$\end{tabular} & \cmark & \cmark & \xmark & \xmark & \xmark & \xmark & \xmark & \xmark & \xmark & \xmark & \cmark & \cmark &\\[15pt]\cline{2-14}  &  \begin{tabular}[c]{@{}c@{}} \xmark~(SC)\\ $\chi_\text{ee}^{xz}=-\chi_\text{ee}^{zx}$\end{tabular} & \cmark & \cmark & \xmark & \xmark & \xmark & \xmark & \xmark & \xmark & \xmark & \xmark & \cmark & \cmark &\\[15pt]
\hline\hline  \multirow{3}{*}[-0.5cm]{\begin{tabular}[c]{@{}c@{}c@{}} Bianisotropic\\$\chi_\text{ee}^{xx}$, $\chi_\text{mm}^{yy}$\\ $\chi_\text{em}^{xy}$, $\chi_\text{me}^{yx}$\end{tabular}} & \begin{tabular}[c]{@{}c@{}} \cmark\\ $\chi_\text{em}^{xy}=-\chi_\text{me}^{yx}$\end{tabular} & \cmark & \cmark & \cmark & \cmark & \cmark & \cmark & \cmark & \cmark & \xmark & \xmark & \xmark & \xmark &\\[15pt]\cline{2-14}  & \begin{tabular}[c]{@{}c@{}} \xmark\\ $\chi_\text{em}^{xy}\neq-\chi_\text{me}^{yx}$\end{tabular} & \xmark & \xmark & \xmark & \xmark & \cmark & \cmark & \cmark & \cmark & \xmark & \xmark & \xmark & \xmark &\\[15pt]\cline{2-14}  & \begin{tabular}[c]{@{}c@{}} \xmark~(SC)\\ $\chi_\text{em}^{xy}=\chi_\text{me}^{yx}$\end{tabular} & \xmark & \xmark & \xmark & \xmark & \cmark & \cmark & \cmark & \cmark & \cmark & \cmark & \cmark & \cmark &\\[15pt]
\hline\hline \multirow{6}{*}[-2.3cm]{\begin{tabular}[c]{@{}c@{}c@{}c@{}} Bianisotropic\\$\chi_\text{ee}^{xx}$, $\chi_\text{mm}^{yy}$, $\chi_\text{ee}^{zz}$ \\$\chi_\text{ee}^{xz}$,$\chi_\text{ee}^{zx}$, $\chi_\text{em}^{xy}$, \\$\chi_\text{me}^{yx}$,$\chi_\text{em}^{zy}$, $\chi_\text{me}^{yz}$\end{tabular}} &  \begin{tabular}[c]{@{}c@{}c@{}c@{}} \cmark\\ $\chi_\text{ee}^{xz}=\chi_\text{ee}^{zx}$\\ $\chi_\text{em}^{xy}=-\chi_\text{me}^{yx}$\\ $\chi_\text{em}^{zy}=-\chi_\text{me}^{yz}$\end{tabular} & \cmark & \cmark & \xmark & \xmark & \xmark & \xmark & \cmark & \cmark & \xmark & \xmark & \xmark & \xmark &\\[20pt]\cline{2-14} & \begin{tabular}[c]{@{}c@{}c@{}c@{}} \xmark\\ $\chi_\text{ee}^{xz}\neq\chi_\text{ee}^{zx}$\\ $\chi_\text{em}^{xy}\neq-\chi_\text{me}^{yx}$\\ $\chi_\text{em}^{zy}\neq-\chi_\text{me}^{yz}$\end{tabular} & \xmark & \xmark & \xmark & \xmark & \xmark & \xmark & \xmark & \xmark & \xmark & \xmark & \xmark & \xmark &\\[25pt]\cline{2-14}
 & \begin{tabular}[c]{@{}c@{}c@{}c@{}} \xmark~(SC)\\ $\chi_\text{ee}^{xz}=\chi_\text{ee}^{zx}$\\ $\chi_\text{em}^{xy}=\chi_\text{me}^{yx}$\\ $\chi_\text{em}^{zy}=\pm\chi_\text{me}^{yz}$\end{tabular} & \xmark & \xmark & \xmark & \xmark & \xmark & \xmark & \xmark & \xmark & \cmark & \cmark & \xmark & \xmark &\\[25pt]\cline{2-14}
  & \begin{tabular}[c]{@{}c@{}c@{}c@{}} \xmark~(SC)\\ $\chi_\text{ee}^{xz}=\pm\chi_\text{ee}^{zx}$\\ $\chi_\text{em}^{xy}=\mp\chi_\text{me}^{yx}$\\ $\chi_\text{em}^{zy}=\chi_\text{me}^{yz}$\end{tabular} & \xmark & \xmark & \xmark & \xmark & \xmark & \xmark & \xmark & \xmark & \xmark & \xmark & \xmark & \xmark &\\[25pt]\cline{2-14}
   & \begin{tabular}[c]{@{}c@{}c@{}c@{}} \xmark~(SC)\\ $\chi_\text{ee}^{xz}=-\chi_\text{ee}^{zx}$\\ $\chi_\text{em}^{xy}=-\chi_\text{me}^{yx}$\\ $\chi_\text{em}^{zy}=\pm\chi_\text{me}^{yz}$\end{tabular} & \xmark & \xmark & \cmark & \cmark & \xmark & \xmark & \xmark & \xmark & \xmark & \xmark & \xmark & \xmark &\\[25pt]\cline{2-14}
    & \begin{tabular}[c]{@{}c@{}c@{}c@{}} \xmark~(SC)\\ $\chi_\text{ee}^{xz}=-\chi_\text{ee}^{zx}$\\ $\chi_\text{em}^{xy}=\chi_\text{me}^{yx}$\\ $\chi_\text{em}^{zy}=-\chi_\text{me}^{yz}$\end{tabular} & \xmark & \xmark & \xmark & \xmark & \cmark & \cmark & \xmark & \xmark & \xmark & \xmark & \cmark & \cmark &\\[25pt]
    \hline
\end{tabular}
  \caption{Properties of symmetry and reciprocity for different types of metasurfaces classified following the convention adopted in Fig.~\ref{fig:Tsym}. For each category, the metasurface is either symmetric/reciprocal (\cmark) or asymmetric/nonreciprocal (\xmark). The label ``(SC)'' indicates a special case of nonreciprocity.}
  \label{tab:summary}
\end{table*}

Let us now consider the case where the metasurface is reciprocal, which according to~\eqref{eq:reciprocity}, implies that the following conditions are simultaneously satisfied $\chi_\text{ee}^{xz}=\chi_\text{ee}^{zx}$, $\chi_\text{em}^{xy}=-\chi_\text{me}^{yx}$ and $\chi_\text{em}^{zy}=-\chi_\text{me}^{yz}$. As explained above, this type of metasurface is completely asymmetric and thus corresponds to the diagrammatic representation of Fig.~\ref{fig:biani_rec}.

Such asymmetric angular scattering properties may, for instance, be achieved with the unit cell design of Fig.~\ref{fig:All}d, where the horizontal rods have a length of 170~nm and 100~nm, and the vertical rods have a length of 75~nm and 45~nm. The resulting amplitude and phase of the scattering parameters are respectively plotted in Figs.~\ref{fig:All}h and~\ref{fig:All}l where the expected angular asymmetric scattering behavior is indeed retrieved. Note that this unit cell has been essentially designed by mixing together the unit cells of Figs.~\ref{fig:All}b and~\ref{fig:All}c.

In the case where the metasurface is nonreciprocal and that all conditions provided above are not satisfied, then both reflection and transmission coefficients are nonreciprocal in addition of being asymmetric. This leads to the diagrammatic representation of Fig.~\ref{fig:biani_NR}.

Due to the complexity of this type of metasurfaces, there are several different ``special cases'' of nonreciprocity that may also occur. We do not discuss them here since they would be particularly difficult to implement in practice. Nevertheless, we have considered some of these special cases and have included their corresponding angular scattering properties in the summary Table~\ref{tab:summary}.

Note that in this table, the susceptibilities associated with the 4 types of metasurfaces are only indicative. Indeed, as said above, it is impossible to make the difference between a birefringent metasurface whose only nonzero susceptibilities are $\chi_\text{ee}^{xx}$ and $\chi_\text{mm}^{yy}$ and a metasurface whose nonzero susceptibilities are $\chi_\text{ee}^{xx}$, $\chi_\text{mm}^{yy}$ and $\chi_\text{ee}^{zz}$ since $\chi_\text{ee}^{zz}$ is not related to an asymmetric function of $\theta$.

\section{Relations Between Scattering Symmetries and Structural Symmetries}
\label{sec:sym}

A close inspection of the scattering particles in Fig.~\ref{fig:All} suggests the possibility to relate their structural symmetries to their corresponding angular scattering properties. In what follows, we thus discuss the relationships between the angular scattering symmetries and the structural symmetries of different scattering particles and deduce some general rules that may be useful to design metasurfaces with specific angular scattering properties.

In the forthcoming discussion, we assume that the origin of the coordinate system lies in the center of the scattering particles. We also consider that the scattering particles exhibit a plane reflection symmetry with respect to the $xz$-plane such as those in Fig.~\ref{fig:All}. The more general case where the scattering particles do not exhibit an $xz$-plane reflection symmetry will be the topic of a future work.

Let us first consider the case where the metasurfaces are reciprocal, which leads to the 4 different types of scattering behavior presented in Fig.~\ref{fig:All}. From this figure, we observe that there are 3 possible types of symmetries~\cite{cornwell1997group}: a $180^\circ$-rotation symmetry around the $y$-axis ($C_2$), a reflection symmetry through the $x$-axis ($\sigma_x$) and a reflection symmetry through the $z$-axis ($\sigma_z$). Accordingly, we report in Table~\ref{tab:sym} the symmetries of the reflection and transmission coefficients as well as those of the corresponding scattering particles for these 4 types of reciprocal metasurfaces.
\begin{table}[h!]
\centering
\begin{tabular}{c||c|c||cN}
Type & Reflection & Transmission & Structure &\\
    \hline\hline
    \begin{tabular}[c]{@{}c@{}} Birefringent\\$\chi_\text{ee}^{xx}$, $\chi_\text{mm}^{yy}$\end{tabular} & $C_2\sigma_{z}$ & $C_2\sigma_{z}$ &  \begin{tabular}[c]{@{}c@{}} $C_2\sigma_{z}$\\(or $\sigma_{x}$)\end{tabular}&\\[15pt]
    \hline
    \begin{tabular}[c]{@{}c@{}c@{}} Anisotropic\\$\chi_\text{ee}^{xx}$, $\chi_\text{mm}^{yy}$\\ $\chi_\text{ee}^{xz}$, $\chi_\text{ee}^{zx}$, $\chi_\text{ee}^{zz}$\end{tabular} & $C_2\sigma_{z}$ & $C_2$ & $C_2$ &\\[23pt]
    \hline
    \begin{tabular}[c]{@{}c@{}c@{}} Bianisotropic\\$\chi_\text{ee}^{xx}$, $\chi_\text{mm}^{yy}$\\ $\chi_\text{em}^{xy}$, $\chi_\text{me}^{yx}$\end{tabular}& $\sigma_{z}$ & $C_2\sigma_{z}$ & $\sigma_{z}$ &\\[23pt]
    \hline
    \begin{tabular}[c]{@{}c@{}c@{}c@{}} Bianisotropic\\$\chi_\text{ee}^{xx}$, $\chi_\text{mm}^{yy}$, $\chi_\text{ee}^{zz}$ \\$\chi_\text{ee}^{xz}$,$\chi_\text{ee}^{zx}$, $\chi_\text{em}^{xy}$, \\$\chi_\text{me}^{yx}$,$\chi_\text{em}^{zy}$, $\chi_\text{me}^{yz}$\end{tabular}& $\sigma_{z}$ & $C_2$ & $-$ &\\[30pt]
    \hline
\end{tabular}
  \caption{Symmetry relationships between angular scattering and unit cell structure for the 4 types of reciprocal metasurfaces. $C_2$ refers to a 2-fold ($180^\circ$) rotation symmetry around the $y$-axis, while $\sigma_{z}$ and $\sigma_{x}$ refer to reflection symmetries through the $z$-axis and the $x$-axis, respectively.}
  \label{tab:sym}
\end{table}

The first important deduction is that reciprocity necessarily implies that the reflection coefficients have a $\sigma_z$ symmetry, while the transmission coefficients have a $C_2$ symmetry. Note that this statement is valid irrespective of the geometry of the scattering particles. The second deduction is that the angular scattering properties of a given reciprocal metasurface always exhibit the same symmetries as those of its scattering particles.

Accordingly, if the scattering particles do not exhibit any structural symmetry, then the resulting metasurface corresponds to a bianisotropic structure with both normal and tangential polarization densities and its angular scattering response does not present any symmetry besides those imposed by reciprocity (last row in Table~\ref{tab:sym}). If the scattering particles are \emph{only} $\sigma_z$ symmetric, then the metasurface is bianisotropic with only tangential polarizations. If the scattering particles are \emph{only} $C_2$ symmetric, then the metasurface is anisotropic with both normal and tangential polarizations. Finally, if the scattering particles are both $C_2$ and $\sigma_z$ symmetric, then the metasurface is birefringent, i.e. like that shown in Figs.~\ref{fig:All}e and~\ref{fig:All}i.

At this point, it should emphasized that the three possible symmetries ($\sigma_x$, $\sigma_z$ and $C_2$) are connected to each other such that if a structure (or an angular scattering diagram as those in Fig.~\ref{fig:All}) exhibits both $C_2$ and $\sigma_z$ symmetries, then it necessarily also exhibits a $\sigma_x$ symmetry due to the fundamental property that $C_2\cdot\sigma_z=\sigma_x$~\cite{cornwell1997group}. For instance, the scattering particle in Fig.~\ref{fig:All}a is $C_2$ and $\sigma_z$ symmetric and it is also de facto $\sigma_x$ symmetric. However, the equality $C_2\cdot\sigma_z=\sigma_x$ may be confusing in some cases. Indeed, take for instance the scattering particle in Fig.~\ref{fig:All}b and flip the bottom L-shaped structure by $180^\circ$ around the $z$-axis such that it corresponds to the mirror reflection of the top L-shaped structure. Such a scattering particle would neither be $\sigma_z$ symmetric nor $C_2$ symmetric but it would be $\sigma_x$ symmetric and would thus exhibit an angular scattering response that is equivalent to that of a birefringent metasurface.

Besides the classification of Tables~\ref{tab:summary} and~\ref{tab:sym} in terms of metasurface types, we can assert the following relationships between structural symmetries and presence of susceptibilities: 1) a scattering particle with full structural symmetry ($C_2\sigma_z$ or $\sigma_x$) corresponds to an effective zero-thickness material which exhibits nonzero $\chi_\text{ee}^{xx}$ and $\chi_\text{mm}^{yy}$ components (and probably also $\chi_\text{ee}^{zz}$ to a lesser extent), 2) a scattering particle with only $C_2$ symmetry  exhibits \emph{at least} nonzero $\chi_\text{ee}^{xz}$ and $\chi_\text{ee}^{zx}$ components, 3) a scattering particle with only $\sigma_z$ symmetry is related to the presence of \emph{at least} $\chi_\text{em}^{xy}$ and $\chi_\text{me}^{yx}$, 4) a scattering particle without any structural symmetry must \emph{at least} exhibit the following susceptibility components: $\chi_\text{ee}^{xz}$, $\chi_\text{ee}^{zx}$, $\chi_\text{em}^{xy}$ and $\chi_\text{me}^{yx}$.

So far, we have discussed the symmetries associated with reciprocal metasurfaces. Accordingly, we have presented, in Table~\ref{tab:sym}, the 4 main types of reciprocal metasurfaces and the corresponding 4 possible combinations of scattering particle symmetries. A nonreciprocal metasurface must somehow be able to break the reciprocal angular scattering symmetries, i.e. $\sigma_z$ symmetry for the reflection and $C_2$ symmetry for the transmission. This is obviously impossible to achieve simply by controlling the geometrical properties of the particles but rather requires the introduction of either: a time-odd bias, a time-varying modulation or some form of nonlinear interactions.

\section{Conclusion}
\label{sec:concl}

In this paper, we have shed some light on the role played by both tangential and normal polarization densities on the angular scattering properties of bianisotropic metasurfaces. To do so, we have considered different types of metasurfaces and have investigated their angular scattering responses in terms of their susceptibilities. We have shown that the angular scattering properties of a metasurface may generally be classified according to 12 different categories composed of 4 types of reciprocal/nonreciprocal scattering and 8 types of symmetrical/asymmetrical scattering. Finally, we have deduced relationships between the structural symmetries of the metasurface scattering particles and the corresponding symmetries of their angular scattering response. This may prove very useful to design metasurface scattering particles that achieve asymmetric angular scattering.

\section*{Acknowledgements}

We gratefully acknowledge funding from the European Research Council (ERC-2015-AdG-695206 Nanofactory).


\bibliographystyle{myIEEEtran}
\bibliography{NewLib}

\end{document}